\documentclass[twocolumns,conference,10pt]{IEEEtran}
\usepackage{epsfig, eucal, amssymb}
\usepackage[cmex10]{amsmath}
\usepackage{subfigure}
\ifCLASSINFOpdf
\else
\fi

\begin{document}
\newtheorem{ach}{Achievability}
\newtheorem{con}{Converse}
\newtheorem{definition}{Definition}
\newtheorem{theorem}{Theorem}
\newtheorem{lemma}{Lemma}
\newtheorem{example}{Example}
\newtheorem{cor}{Corollary}
\newtheorem{prop}{Proposition}
\newtheorem{conjecture}{Conjecture}
\newtheorem{remark}{Remark}
\title{Optimal Multiplexing Gain of $K$-user Line-of-Sight Interference Channels with Polarization}
\author{\IEEEauthorblockN{Sung ho Chae, Sang Won Choi, and Sae-Young
Chung} \IEEEauthorblockA{School of EECS, KAIST, 335 Gwahangno,
Yuseong-gu, Daejeon, 305-701, South Korea} {Email:
shchae@kaist.ac.kr,  ace1905@kaist.ac.kr, and
sychung@ee.kaist.ac.kr}
} \maketitle

\begin{abstract}
We consider the multiplexing gain (MUXG) of the fully connected
$K$-user line-of-sight (LOS) interference channels (ICs). A polarimetric
antenna composed of 3 orthogonal electric dipoles and
$3$ orthogonal magnetic dipoles is considered where all 6 dipoles are co-located. In case of $K$-user IC with single polarization, the
maximum achievable MUXG is $K$ regardless of the
number of transmit and receive antennas because of the key-hole
effect. With polarization, a trivial upper bound on the MUXG is $2K$.
We propose a zero forcing (ZF) scheme for the $K$-user LOS IC, where each user uses one or more polarimetric antennas. 
By using the proposed ZF scheme, we find minimal antenna configurations that achieve this
bound for $K\leq 5$. For $K>5$, we show that the optimal MUXG of $2K$ is
achieved with $M=\lceil\frac{K+1}{6}\rceil$ polarimetric antennas
at each user.

\end{abstract}

\IEEEpeerreviewmaketitle

\section{Introduction}
In multi-user communications, resources such as time, frequency, and
antennas need to be allocated to each user properly. 
One simple way is to allocate resources orthogonally. However, it can
be very inefficient. To improve the performance further
resources can be shared, which becomes more important as the number ($K$) of users
increases. Studying the interference channel (IC) can give us insights 
on how to manage interference better.

Exact capacity characterization of the Gaussian IC is unknown in
general. However, approximate characterization by finding the
optimal multiplexing gain has been found for many cases recently.
For example, zero forcing has been shown to be efficient for
$K$-user multiple-input multiple-output (MIMO) Gaussian IC~\cite{Jafar07}.
Recently, interference alignment (IA)~\cite{Ali06}-\cite{Etkin09} when $K \geq
3$ has been shown to achieve the optimal degrees of freedom (DOF) of $\frac{K}{2}$ for time-varying channels. 
Furthermore, the IA scheme for line-of-sight (LOS) channels has
been developed in~\cite{Grokop08}.

MIMO is in general helpful for improving the multiplexing gain.
However, in LOS channels, e.g., roof-top antennas,
it is more challenging to achieve a higher multiplexing gain since regardless of the number of 
antennas the multiplexing gain can only be one in a point-to-point setup due to the key hole effect.
Polarization can increase the DOF in this situation. It can provide two-fold increase in
the DOF in a LOS environment. 
In~\cite{nature}, the authors showed that
up to $6$ DOF can be obtained by using a single polarimetric antenna in a scattering environment. 
In~\cite{poon}, the multiplicative gain in DOF was studied
for different array geometry when polarimetric antennas are used. The results showed that
the multiplicative gain from polarization depends on the array
geometry and the scattering condition.

In this paper, we focus on the MUXG of the fully connected $K$-user
LOS IC. We propose a new interference cancellation scheme using the
polarimetric antenna studied in~\cite{nature},~\cite{poon}. Due to the key hole effect,
the total MUXG in the $K$-user LOS IC
is upper bounded by $K$ with single polarization. If we use
polarimetric antennas at each user, we observe that a trivial upper bound on the DOF is
now $2K$. In this paper, we find minimal antenna configurations to achieve
this trivial upper bound. For $K\leq 5$, we show
some antenna configurations to achieve the optimal MUXG of $2K$. 
We also show that for $K>5$, we need $\lceil\frac{K+1}{6}\rceil$
polarimetric antennas per node to achieve the optimal MUXG.

\begin{figure}[!t]
\centering
\includegraphics[width=1.6 in]{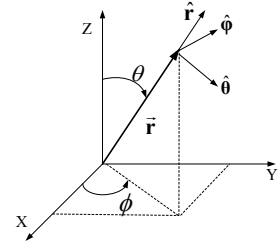}
\caption{Spherical coordinate system and corresponding unit vectors}
\label{fig 1}
\end{figure}

\section{Channel Model}
We consider a polarimetric antenna composed of $3$
orthogonal electric dipoles and $3$ orthogonal magnetic dipoles that
are all co-located as
in ~\cite{nature},~\cite{poon}. Note that since the space is $3$
dimensional we can try to have up to $3$ electric and $3$ magnetic
dipoles at the same location to construct a polarimetric antenna. We
interpret this polarimetric antenna as a vector point source having 6 elements.

We will first see the radiated electric fields from each electric
dipole and magnetic dipole. For convenience, a spherical coordinate
is considered throughout this paper as in Fig. \ref{fig 1}. Consider
electric dipoles oriented along the $x$-, $y$- and $z$-axis, which
are positioned symmetrically at the origin. We assume all the
electric dipoles are half-wave dipoles, and their maximum electric
currents are $I_{e1}$, $I_{e2}$, and $I_{e3}$, respectively. Then,
the radiated electric fields from each electric dipole measured at a
point $\vec{r}$ in the far-field are given by ~\cite{balanis}
\begin{align*}
E_1=E^{x}_e&\simeq
j\sqrt{\frac{\mu}{\epsilon}}\frac{I_{e1}e^{-jkr}}{2\pi
r}[-\cos\theta\cos\phi\mathbf{\hat{\theta}}+\sin\phi\mathbf{\hat{\phi}}]\\
&=C_{e1}\frac{e^{-jkr}}{r}\mathbf{a}_1(\theta,\phi),\\
E_2=E^{y}_e&\simeq
j\sqrt{\frac{\mu}{\epsilon}}\frac{I_{e2}e^{-jkr}}{2\pi
r}[-\cos\theta\sin\phi\hat{\theta}-\cos\phi\hat{\phi}]\\
&=C_{e2}\frac{e^{-jkr}}{r}\mathbf{a}_2(\theta,\phi),
\end{align*}
and \begin{align*} E_3=E^{z}_e&\simeq
j\sqrt{\frac{\mu}{\epsilon}}\frac{I_{e3}e^{-jkr}}{2\pi r}[\sin\theta
\bold{\hat{\theta}}]=C_{e3}\frac{e^{-jkr}}{r}\mathbf{a}_3(\theta,\phi),
\end{align*}
respectively, where $k$ denotes the wave number,
$\sqrt{\frac{\mu}{\epsilon}}\approx377\Omega$ is the impedance of
free space, $r=\mid \vec{r} \mid$ is the distance between the
observation point and the origin, and
$C_{ei}=j\sqrt{\frac{\mu}{\epsilon}}\frac{I_{ei}}{2\pi}$, $\forall
i=1,2,3$.

Similarly, the radiated electric fields from magnetic dipoles with
the maximum magnetic current $I_{m4}$, $I_{m5}$, and $I_{m6}$ are
given by
\begin{align*}
E_4=E^{x}_m&\simeq j\sqrt{\frac{\mu}{\epsilon}}\frac{I_{m4}e^{-jkr}}{2\pi r}[\sin\phi\mathbf{\hat{\theta}}+\cos\theta\cos\phi\mathbf{\hat{\phi}}]\\
&=C_{m4}\frac{e^{-jkr}}{r}\mathbf{a}_4(\theta,\phi),\\
E_5=E^{y}_m&\simeq
j\sqrt{\frac{\mu}{\epsilon}}\frac{I_{m5}e^{-jkr}}{2\pi
r}[-\cos\phi\hat{\theta}+\cos\theta\sin\phi\hat{\phi}]\\
&=C_{m5}\frac{e^{-jkr}}{r}\mathbf{a}_5(\theta,\phi),
\end{align*}
and \begin{align*} E_6=E^{z}_m&\simeq
j\sqrt{\frac{\mu}{\epsilon}}\frac{I_{m6}e^{-jkr}}{2\pi
r}[-\sin\theta\hat{\phi}]=C_{m6}\frac{e^{-jkr}}{r}\mathbf{a}_6(\theta,\phi)
\end{align*}
from duality theorem, where
$C_{mi}=j\sqrt{\frac{\mu}{\epsilon}}\frac{I_{mi}}{2\pi}$, $\forall
i=4,5,6$. However, since there are no known magnetic currents in
nature, magnetic current is purely a mathematical notion used to explain the
motion of magnetic charges creating magnetic current, when compared
to their dual quantities of moving electric charges giving rise to
electric current~\cite{time}. In fact, we can generate the
equivalent radiation pattern by a closed loop of electric current.
The equivalent relationship between the magnetic current $I_m$ and
the electric current of the loop antenna $I_l$ is given by  $I_m
\lambda /2=j\pi a \left(2\pi f \right)^2\mu_0 I_l$, where $a$ is the
radius of the loop antenna, $f$ is the operating frequency, and
$\mu_0$ is permeability of free space. Note that the maximum
electric (magnetic) currents of each dipole are determined by their
transmit powers.

We now consider the fully connected $K$-user LOS IC with
polarimetric transmit and receive antennas. Transmitter $i$ tries to
communicate with receiver $i$ that is getting interference from all the other transmitters, where $\forall
i=1,~2,~\cdots,~K$, and each transmitter and receiver uses $M$
polarimetric antennas. In addition, we assume all channel
coefficients are fixed during the communication duration (time
invariant) and known to all transmitters and receivers. Then, the
input and output relationship is given by
\begin{equation*}
\mathbf{Y}^{[j]}=\sum^{K}_{i=1}\mathbf{H}^{[ij]}\mathbf{X}^{[i]}+\mathbf{N}^{[i]}
\end{equation*}
where $\mathbf{X}^{[i]}$ is the $6M\times 1$ signal vector at the
transmitter $i$, $\mathbf{H}^{[ij]}$ is the $6M\times 6M$
polarization matrix from the transmitter $i$ to the receiver $j$,
and $\mathbf{Y}^{[j]}$ is the $6M\times 1$ the received signal vector at
the receiver $j$. The noise vector $\mathbf{N}^{[j]}$ is the
additive white Gaussian with zero mean and covariance of
$\mathbf{I}_{6M}$, where $\mathbf{I}_{N}$
denotes the identity matrix of size $N\times N$.

In this paper, we assume that all the transmitters and receivers are
located in the azimuth plane $\left(\theta=\frac{\pi}{2}\right)$ only.
Note that this assumption holds roughly for typical LOS channels found in
cellular networks.

In order to see the characteristics of polarization matrix, fix $i$
and $j$, and set $M=1$. Note that the $6 \times 6$ polarization
matrix $\mathbf{H}^{[ij]}$ shows the voltage response of each
component of the receiver $j$ due to the incident wave radiated by
each component of the transmitter $i$. Let $a^{[ij]}$ denote the
attenuation along the LOS path between transmitter $i$ and receiver
$j$, and let $\phi_{ij}$ denote the angle between the transmitter $i$
and the receiver $j$. Then, the channel matrix $\mathbf{H}^{[ij]}$
is given by
\begin{align*}
&\mathbf{H}^{[ij]}=a^{[ij]} e^{-jkr_{[ij]}}\left[a_{\hat{\theta}}^T
a_{\hat{\theta}}+a_{\hat{\phi}}^T a_{\hat{\phi}}\right]
\end{align*}
where $r_{[ij]}$ denotes the distance between the transmitter $i$
and the receiver $j$, and the operator $(\cdot)^T$ denotes the
transpose of a matrix. In addition, we have $[\begin{array}{rrrrrrr}
\mathbf{a}_1(\frac{\pi}{2},\phi_{ij})
&\mathbf{a}_2(\frac{\pi}{2},\phi_{ij})&\cdots&\mathbf{a}_6(\frac{\pi}{2},\phi_{ij})\end{array}]=a_{\hat{\theta}}\hat{\theta}+a_{\hat{\phi}}\hat{\phi}$
   ~where $a_{\hat{\theta}}=[\begin{array}{rrrrrrr} \sin(\phi_{ij})
&-\cos(\phi_{ij}) &0 &0 &0 &-1\end{array} $] and
$a_{\hat{\phi_{ij}}}= [\begin{array}{rrrrrr} 0 &0
&1&\sin(\phi_{ij})&-\cos(\phi_{ij})&0
\end{array}]$.
Note that $a_{\hat{\theta}}$ and $a_{\hat{\phi}}$ represent the
relative magnitude of electric fields polarized in $\hat{\theta}$
and $\hat{\phi}$ directions, respectively, which are radiated by each
component of the transmit polarimetric antenna with the same amount of
current. We see that electric dipoles oriented along $x$- and
$y$-axis and magnetic dipole oriented along $z$-axis transmit
horizontally polarized electric fields while the others transmit
vertically polarized electric fields. Clearly, the rank of this
matrix is $2$ even though we use $6$ different dipole components
since there are only two degrees of freedom for a fixed
propagation direction. Moreover, the channel matrix is symmetric due
to the reciprocity of transmitter and receiver.

We follow the conventional definition of achievable rate and
capacity region, which is omitted in this paper. The multiplexing
gain $\Gamma$ \cite{Zheng03} of the $K$-user LOS IC is defined as
\begin{align*}
\Gamma=\underset{\mbox{\small SNR} \rightarrow \infty}{\lim}
&\frac{R_{+}(\mbox{SNR})}{\log(\mbox{SNR})},
\end{align*}
where $R_{+} (\mbox{SNR})$ is a sum rate at signal-to-noise ratio
(SNR), where $\mbox{SNR}$ is defined as the ratio of the total power across all
transmitters and the noise variance at each receiver.

\begin{figure}[!t]
\centering
\includegraphics[scale=0.74]{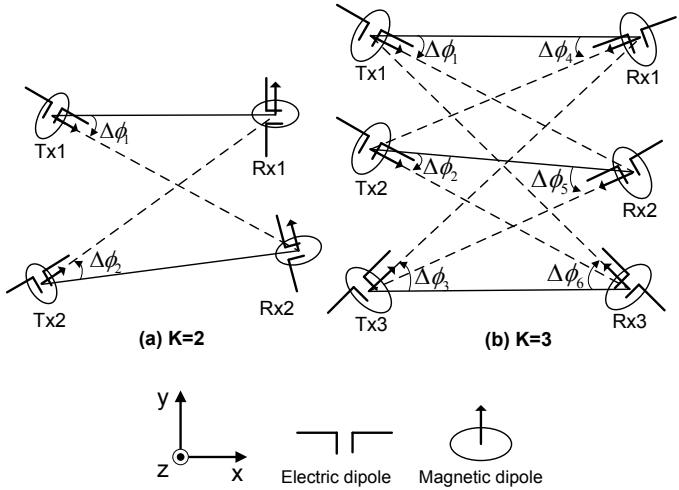}
\caption{Optimal antenna placement for K=2 and K=3 cases. Each
transmitter and receiver is aligned along the cross-link direction
which each of them wants to remove.} \label{fig 2}
\end{figure}

In this paper, we focus on the optimal MUXG for $K$-user LOS
interference channel using polarimetric antennas. Our goal is to
find the minimum antenna configurations that achieve the optimal
MUXG of $2K$. Note that without polarimetric antennas we can only
achieve the MUXG of $K$ instead of $2K$ in a LOS environment even if
we allow to use multiple polarimetric antennas at each node. We propose two different
scenarios, optimal antenna placement and fixed antenna placement. We
will show the minimum antenna configuration under these scenarios,
respectively. Optimal antenna placement means optimally adjusting
the orientation of each dipole component physically. Such manual
adjustment of dipole direction may be feasible if node locations
are exactly known and do not change. In our second scenario, fixed
antenna placement, all antenna directions are fixed
independent of node locations. In this scenario, we may need more dipoles
than in optimal antenna placement scenario. 

\section{Optimal Antenna Placement}

In this section, we find minimal antenna configurations need to achieve
the trivial upper bound of $2K$ on the DOF in the optimal antenna placement
scenario. Note that we need at least 2 dipoles per user to achieve the goal.
Consider the example of $2$-user ($K=2$) case as
shown in Fig. \ref{fig 2}-(a). Assume that all the transmitters and
receivers have one electric and one magnetic
dipoles that are co-located such that they are oriented along the direction of its direct
link initially. In this case, the electric dipole is used to generate
vertically polarized electric field, while the magnetic dipole generates
horizontally polarized electric field. The key idea is based on the
fact that the radiated electric field is always perpendicular to the
direction of propagation, and the its magnitude is proportional to
the projection of the antenna orientation to the field direction.
Therefore, if the dipoles of transmitters $1$ and $2$ are physically
rotated by $\Delta\phi_1\left(=\phi_{12}-\phi_{11}\right)$ and
$\Delta\phi_2\left(=\phi_{21}-\phi_{22}\right)$, respectively where
$\phi_{ij}$ denotes the angle between transmitter $i$ and receiver
$j$, the transmitters do not radiate interfering signal to
non-desired receivers. Since the interference is already canceled by
each transmitter, the dipoles of receivers are rotated by $90$
degrees to get the maximum antenna gain. It is clear that the role
of the transmitters and the receivers can be swapped. Then, after
adjusting the orientation of each dipole properly, we have the
following channel matrices
\begin{align*}
&\mathbf{H}^{[11]}=a^{[11]}e^{-jkr_{[11]}}\left[\begin{array}{cc}
\sin(\Delta\phi_1)&0\\
0&\sin(\Delta\phi_1)
\end{array}
\right]
\end{align*}
and
\begin{align*}
&\mathbf{H}^{[22]}=a^{[22]}e^{-jkr_{[22]}}\left[\begin{array}{cc}
\sin(\Delta\phi_2)&0\\
0&\sin(\Delta\phi_2)
\end{array}
\right]
\end{align*}
while both $\mathbf{H}^{[12]}$ and $\mathbf{H}^{[21]}$ become null
matrices. We see that if $\phi_{11}\neq \phi_{12}$ and $\phi_{22}\neq \phi_{21}$,
$2$ DOF per each user can be achieved. One can see that there can be
other proper antenna configurations for achieving the same MUXG. 
Suppose the first user uses one electric
and one magnetic dipole that are co-located and oriented along $z$- and $x$-axis,
respectively while the second user uses the same dipole set as the
previous case. In this case, the channel matrix is given by
\begin{align*}
&\mathbf{H}^{[11]}=a^{[11]}e^{-jkr_{[11]}}\left[\begin{array}{cc}
1&0\\
0&\sin(\Delta\phi_1)
\end{array}
\right] \end{align*} and
\begin{align*}
&\mathbf{H}^{[22]}=a^{[22]}e^{-jkr_{[22]}}\left[\begin{array}{cc}
\sin(\Delta\phi_2)&0\\
0&\sin(\Delta\phi_2)
\end{array}
\right]
\end{align*}
and the optimal MUXG is also achieved. However, if transmitter
and receiver use electric dipoles only, we cannot remove the
interference from the other user.

We can easily extend this scheme to $K=3$ by using same
set of dipoles. However, since there are $2$ interference links per each
user in this case, the dipoles of both transmitters and receivers
need to be rotated in order to remove all the interferences. One
proper antenna configuration for $K=3$ is depicted in Fig. \ref{fig
2}-(b), and the following channel matrices can be easily obtained
for this case:
\begin{align*}
&\mathbf{H}^{[ii]}=a^{[ii]}e^{-jkr_{[ii]}}\left[\begin{array}{cc}
\lambda^{[ii]}&0\\
0&\lambda^{[ii]}
\end{array}
\right], \quad \forall i=1,2,3
\end{align*}
where
$\lambda^{[11]}=\sin(\phi_{11}-\phi_{12})\sin(\phi_{11}-\phi_{21})$,
$\lambda^{[22]}=\sin(\phi_{22}-\phi_{21})\sin(\phi_{22}-\phi_{12})$,
and
$\lambda^{[33]}=\sin(\phi_{33}-\phi_{31})\sin(\phi_{33}-\phi_{13})$
while the other channel matrices are null. Similar to $K=2$ case, we
achieve the optimal MUXG of $2K$ by rotating dipoles properly.
Observe that unlike $2$-user case, we cannot use electric or
magnetic dipole oriented along $z$-axis since their radiation
patterns are omni-directional in the azimuth plane. In summary, we have
the following proposition for optimal placement.

\begin{prop}
For $K\leq3$ with optimal antenna placement, the following antenna
configuration at each user achieves the optimal MUXG of $2K$.
\begin{itemize}
\item ($e_x$ or $e_y$) + ($m_y$ or $m_y$)
\end{itemize}
where $e$ and $m$ denote electric and magnetic dipoles respectively,
and the subscript denotes the orientation of each dipole. 
\end{prop}

When $K$
is greater than $3$, we can select a proper subset of $3$ users to
achieve up to $6$ degrees of freedom.

The proposed zero forcing scheme is simple and intuitive,
but we have to adjust dipole orientation physically whenever the
location of either a transmitter and receiver is changed. Therefore,
this scheme can only be feasible if there is no mobility. Furthermore, we
cannot use this scheme when $K>3$. In Section IV, we
explain how to solve these problems by using more polarimetric antenna elements.

\section{Fixed Antenna Placement: $M=1$ Case}
It is easy to see that the electric (magnetic) dipoles oriented
along $x$- and $y$-axis together can create the electric field
generated from an arbitrary rotated dipole in the azimuth plane. Thus,
instead of rotating dipoles physically, we can obtain the same
interference canceling effect by using more polarimetric antenna elements whose
orientation is fixed. In this section, we focus on the case where
$M=1$.

The optimal antenna placement scheme depicted in Fig. \ref{fig
2}-(b) can be replaced by simple ZF beam-forming scheme. The key
idea is that both transmitter and receiver can eliminate a certain
number of its interferences under our scheme. From now on, we will
explain the proposed ZF beam-forming scheme for transmitter and
receiver, respectively.

\begin{figure}[!t]
\centering
\includegraphics[scale=0.85]{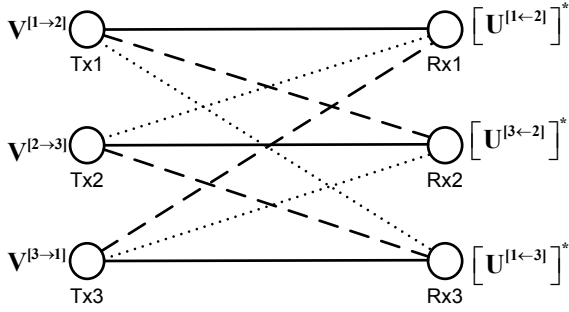}
\caption{Fixed antenna placement with single polarimetric antenna
for $K=3$. The ZF beam-forming matrices for each transmitter and
receiver are depicted. Dashed line($--$) denotes the interference
links nulled by receivers while dotted($\cdots$) line represents the
interference links nulled by transmitters.} \label{fig 3}
\end{figure}

Consider the $3$-user case first. One can easily see that $4$ dipole
components are at least required, $2$ for vertical polarization and
$2$ for horizontal polarization. We assume each user activates $2$
electric and $2$ magnetic dipoles oriented along $x$- and $y$-axis. Then,
each effective channel becomes the $4 \times 4$ matrix whose rank is
$2$. Assume the transmitter $i$ tries to null out the interference
to the receiver $j$. Then, the channel matrix $\mathbf{H}^{[ij]}$ is
given by
\begin{align*}
&\mathbf{H}^{[ij]}=a^{[ij]}e^{-jkr_{[ij]}}\left[\begin{array}{cc}
\mathbf{A}^{[ij]}& \mathbf{O}_{4\times4}\\
\mathbf{O}_{4\times4} & \mathbf{A}^{[ij]}
\end{array}
\right]
\end{align*}
where
\begin{align*}
&\mathbf{A}^{[ij]}= \left[\begin{array}{cc} \sin^2(\phi_{ij})&
-\sin(\phi_{ij})\cos(\phi_{ij})\\
-\sin(\phi_{ij})\cos(\phi_{ij})& \cos^2(\phi_{ij})
\end{array}
\right]
\end{align*}
is $2\times2$ matrix whose rank is $1$, and $\mathbf{O}_{4\times4}$
denotes the $4\times 4$ zero matrix. We can easily find one proper
ZF beam-forming matrix for the transmitter $i$ given by
\begin{align*}
&\mathbf{V}^{[i\rightarrow j]}= \left[\begin{array}{cc}
\cos(\phi_{ij})&
0\\
\sin(\phi_{ij})&0\\
0& \cos(\phi_{ij})\\
0& \sin(\phi_{ij})\\
\end{array}
\right].
\end{align*}
Similarly, since the channel is symmetric, if receiver $i$ intends to
cancel the interference from the transmitter $k$ where $k\neq i$ the
beam-forming matrix for the receiver $i$ can be easily obtained as
\begin{align*}
&\mathbf{U}^{[k\leftarrow i]}= \left[\begin{array}{cc}
\cos(\phi_{ki})&
0\\
\sin(\phi_{ki})&0\\
0& \cos(\phi_{ki})\\
0& \sin(\phi_{ki})\\
\end{array}
\right].
\end{align*}
Therefore, after applying the ZF beam-forming for user $i$, we have
the equivalent parallel channel given by
\begin{align*}
\mathbf{\Lambda}^{[ii]}=&\left[\mathbf{U}^{[k\leftarrow i]}\right]
^{\ast}\mathbf{H}^{[ii]}\mathbf{V}^{[i\rightarrow j]}\\
=&a^{[ii]}e^{-jkr_{[ii]}}\left[\begin{array}{cc}
\lambda^{[ii]} & 0\\
0& \lambda^{[ii]}\\
\end{array}
\right]
\end{align*}
where
$\lambda^{[ii]}=\sin(\phi_{ii}-\phi_{ij})\sin(\phi_{ii}-\phi_{ki})$,
and the operator $(\cdot)^{\ast}$ denotes the complex conjugate
transpose of a matrix. The detailed ZF beam-forming for each
transmitter and receiver are summarized in Fig. \ref{fig 3}. Note
that $\mathbf{\Lambda}^{[ii]}$ has two non-zero diagonal elements
unless $\sin(\phi_{ii}-\phi_{ij})$ or $\sin(\phi_{ii}-\phi_{ki})$ are
zero $\forall i=1,2,3$. Furthermore, one can easily observe
that this ZF beam-forming scheme achieves the exactly same
DOF as in optimal antenna placement scenario.

In addition, it is obvious that the proposed ZF method can be also
applied to $K=2$ case. For $K=2$ case, only transmitters or
receivers need to perform ZF beam-forming because there is only one
interference link per user.

It is worth to mention that we can also use $3$ orthogonal electric
dipoles plus $1$ magnetic dipole oriented along $x$- or $y$-axis, or
$3$ orthogonal magnetic dipoles plus $1$ electric dipole oriented
along $x$- or $y$-axis instead to achieve the optimal MUXG of $6$
since there are $2$ dipole components for each vertical and
horizontal polarization, respectively. However, if we use $3$
orthogonal electric dipoles plus $1$ magnetic dipole oriented along
$z$-axis, or $3$ orthogonal magnetic dipoles plus $1$ electric
dipole oriented along $z$-axis, the optimal MUXG cannot be obtained
under our scheme. Remind that we need to have at least $2$
components for each polarization direction to generate the electric
field in an arbitrary direction.

We showed the optimal MUXG can be achievable using only $4$ dipole
components among $6$. However, as $K$ increases, we cannot remove
all the interference links using only $4$ components. Observe that
each transmitter and receiver is able to eliminate one of its
interferences using the fact that the each polarization channel
matrix has $2$ zero singular values. Therefore, the maximum number
of interference links we can remove is $2K$, but there are $K(K-1)$
cross links in the fully connected $K$-user IC. Hence, we cannot
remove all the interference links when $K>3$ using only $4$ dipole
components.

If we use all the $6$ components of polarimetric antenna at each
user, each channel matrix has $4$ zero singular values, thus each
transmitter and receiver can null out up to $2$ interference links
in a similar way. Suppose the transmitter $i$ intends to null out
the interferences radiated to receivers $j$ and $k$
simultaneously, and let $\mathbf{H}^{[i\rightarrow
jk]}=\left[\begin{array}{cc}\mathbf{H}^{[ij]} &
\mathbf{H}^{[ik]}\end{array}\right]^T$ where $j\neq k$. Since the
$\mathbf{H}^{[i\rightarrow jk]}$ is a $12 \times 6$ matrix and its
rank is $4$, we can easily find one proper ZF beam-forming matrix
for transmitter $i$ whose size is $6 \times 2$ as given by:
\begin{align*}
\mathbf{V}^{[i\rightarrow
jk]}=&\frac{1}{\sqrt{1+\sin^2(\phi_{ij}-\phi_{ik})}}\\
&\cdot\left[\begin{array}{cc}
\cos(\phi_{ij})-\cos(\phi_{ik})&0\\
\sin(\phi_{ij})-\sin(\phi_{ik})&0\\
0&\sin(\phi_{ij}-\phi_{ik})\\
0&\cos(\phi_{ij})-\cos(\phi_{ik})\\
0&\sin(\phi_{ij})-\sin(\phi_{ik})\\
\sin(\phi_{ij}-\phi_{ik})&0
\end{array}
\right]
\end{align*}
In addition, assume receiver $i$ tries to remove the interferences
from transmitters $l$ and $m$ simultaneously, where $i,j,k,l$ and $m$
are all different integers. Then, a proper beam-forming matrix for
receiver $i$ can be easily obtained in a similar manner, and user
$i$ obtains the following parallel channel after applying ZF
beam-forming:
\begin{align*}
\mathbf{\Lambda}^{[ii]}&=\left[\mathbf{U}^{[lm\leftarrow i]}\right]
^{\ast}\mathbf{H}^{[ii]}\mathbf{V}^{[i\rightarrow jk]}\\
=&\frac{a^{[ii]}e^{-jkr_{[ii]}}}{\sqrt{\left(1+\sin^2(\phi_{ij}-\phi_{ik})\right)
\left(1+\sin^2(\phi_{li}-\phi_{mi})\right)}}
\\
&\cdot\left[\begin{array}{cc}
\gamma^{[ii]} & 0\\
0& \gamma^{[ii]}\\
\end{array}
\right]
\end{align*}
where
\begin{align*}
\gamma^{[ii]}=&(\sin(\phi_{ii}-\phi_{ij})-\sin(\phi_{ii}-\phi_{ik})+\sin(\phi_{ij}-\phi_{ik}))\\
       &\cdot(\sin(\phi_{ii}-\phi_{li})-\sin(\phi_{ii}-\phi_{mi})+\sin(\phi_{li}-\phi_{mi}))
\end{align*}

\begin{figure}[!t]
\centering
\includegraphics[scale=0.8]{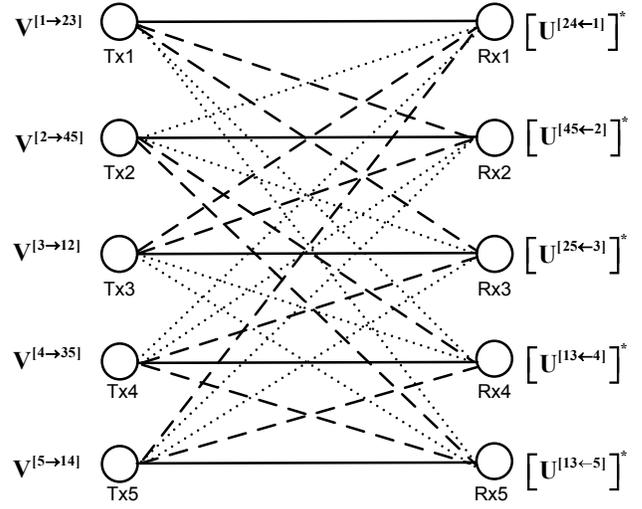}
\caption{Fixed antenna placement with single polarimetric antenna
for $K=5$. The ZF beam-forming matrices for each transmitter and
receiver are depicted. Dashed line$(--)$ denotes the interference
links nulled by receivers while dotted line$(\cdots)$ represents the
interference links nulled by transmitters.} \label{fig4}
\end{figure}

The detailed beam-forming scheme is depicted in Fig. \ref{fig4}. If
every transmitter and receiver null out $2$ interference links, the
maximum number of interfering links we can remove is given by $4K$,
therefore this scheme can be valid for the $K\leq5$ (necessary
condition). Fig. \ref{fig4} shows that the optimal MUXG of $2K$
can be indeed achieved for $K=5$.

In summary, we have the following two propositions for fixed antenna
placement scenario.

\begin{prop}
For $K=2$ or $3$ with fixed antenna placement, the following antenna
configurations at each user achieve the optimal MUXG of $2K$.
\begin{itemize}
\item ($e_x$ and $e_y$) + ($m_y$ and $m_y$)
\item ($e_x$, $e_y$, and $e_z$) + ($m_x$ or $m_y$)
\item ($e_x$ or $e_y$) + ($m_x$, $m_y$, and $m_z$)
\end{itemize}
\end{prop}

\begin{prop}
For $K=4$ or $5$ with fixed antenna placement, using all $6$ dipole
components at each user achieves the optimal MUXG of $2K$.
\end{prop}

In this section, we showed the optimal MUXG can be obtained using
only one polarimetric antenna when $K\leq 5$. However, when $K>5$,
multiple polarimetric antennas ($M>1$) are required.

\section{Fixed Antenna Placement: $M\geq 1$ Case}
In this section, we consider the case where each user uses $M\geq1$
polarimetric antennas. For simplicity, assume that all $6$
components of polarimetric antenna is activated. In this case, the
size of each channel matrix becomes $6M \times 6M$. However, the
rank of this matrix remains at $2$ due to the key-hole effect
because we only consider the LOS environment and far-field
communication. In addition, the geometries of the transmit and
receive array also do not affect the rank of the channel matrix.
Therefore, we see that each transmitter and receiver can remove up
to $3M-1$ interferences connected to itself because the number of
zero singular values is $6M-2$. In addition, we assume any column
vector which lies in the null space of the channel matrix $H^{[ij]}$
is independent of any column vector which spans the null space of
$H^{[kl]}$ except when $i=j$ and $k=l$ since the null space of a
channel matrix is a function of the angle between transmitter and
receiver. Then, the rank of any direct links remains $2$ after
applying our scheme. Using these properties and assumptions, we have
the following proposition.
\\
\begin{prop}
For the $K$-user LOS IC with $M$ polarimetric antennas at each user,
the optimal MUXG of $2K$ is achieved if
$M=\lceil\frac{K+1}{6}\rceil$. \label{Thm}
\end{prop}



\section{Numerical Results}

\begin{figure}[!t]
\centering
\includegraphics[scale=0.63]{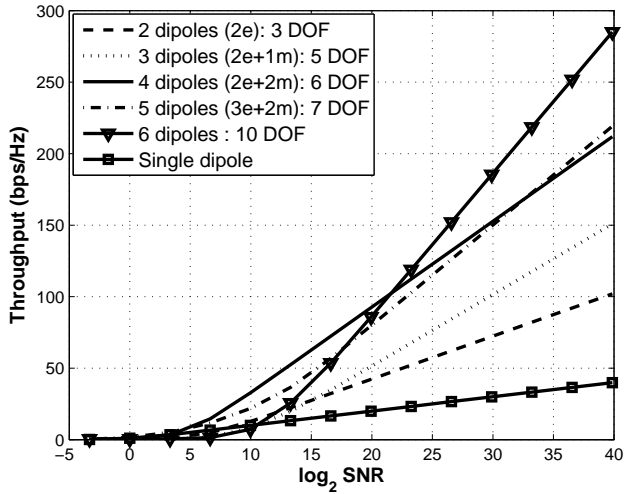}
\caption{Multiplicative gains from polarization for $5$ user LOS
IC.} \label{num}
\end{figure}

In this section, we provide some numerical results demonstrating the
performance of our scheme. We show achievable multiplicative gains
from polarization for $5$-user LOS IC. As can be seen from Fig.~\ref{num}, 
the MUXG increases as we use more dipole components and it reaches the maximum of 10 when we use all 6 dipoles. In
addition, we can observe that using $2$ orthogonal dipoles is not
enough.

%

\section*{Acknowledgment}
This research was supported in part by the ITRC (NIPA-2009-C1090-0902-0005).

\end{document}